\documentstyle[11pt,newpasp,twoside,epsf,epsfig,hyperref]{article}
\hypersetup{pdftitle={Methane Opacities in T Dwarf Atmospheres},
  pdfsubject={Molecular Opacities},
  pdfauthor={Derek Homeier, UGA <derek@physast.uga.edu>}, 
  pdfkeywords={brown dwarfs, molecular data, infrared spectroscopy}
}
\reversemarginpar 
 
\begin{document} 
\title{Methane Opacities in T Dwarf Atmospheres} 
 \author{Derek Homeier} 
 \author{Peter H.~Hauschildt} 

\affil{Department of Physics \& Astronomy and Center for Simulational 
 Physics, University of Georgia, Athens, GA 30602-2451} 
\author{France Allard} 
\affil{Centre de Recherche Astronomique de Lyon, Ecole Normale 
 Sup{\'e}rieure, Lyon, France} 
 
\begin{abstract} 
We present the current status of { PHOENIX} model atmospheres for dwarfs
of the spectral type T, typical for older field brown dwarfs and low-mass
brown dwarfs. The results are based on new predictions of the CH$_4$ 
line opacities from theoretical calculations with the {STDS} software
package, extrapolating to transitions from rotational levels up to
$J\,=\,40$. 
While individual line positions and strengths are reproduced with
moderate to fair accuracy, the cumulative band strength in the region
of the IR methane bands is modelled much better thanks to the 
inclusion of large numbers of faint lines relevant at high
temperatures. 
\end{abstract} 
 
\section{Methane Dwarfs}
The transition between spectral classes L and T, as
$T_{\rm eff}$ drops below $\sim$\,1300\,K, is characterised by a
reduction of the effects of dust absorption and the appearance of 
CH$_4$ absorption bands in the infrared, showing up in the 
coolest brown dwarfs which have only recently been discovered in
substantial numbers in the field. 
These features appearing in the K and, from spectral
type T0, in the H band and growing in strength with decreasing
temperature, are one of the basic characteristics defined in the
classification schemes of Geballe et al.\ (2002) and Burgasser et al.\
(2002) for the T spectral class.  
As one of the dominant opacity sources
in these objects besides H$_2$O and CIA H$_2$, the CH$_4$ molecular
bands are also of great importance for the correct modelling of
ultracool substellar atmospheres and for evolutionary calculations of
old brown dwarfs. 

\section{Line Opacity Predictions}
A major problem in model atmosphere calculations for substellar
objects is the paucity of molecular opacity data for the temperature
regime of brown dwarfs. Available line lists from spectroscopic
databases for Earth and planetary atmospheres, based on 
both experimental data and theoretical predictions, are compiled for
much lower temperatures and thus often lack 
transitions from lower states at higher energies.

Theoretical approaches to model the energy levels and transition
probabilities of CH$_4$ are challenged by the complexity of the 
vibrational systems ({polyads}) of the molecule, due to 
resonances of the four vibrational modes. This leads to eight
vibrational bands with 24 subbands already in the third system 
({Octad}) and 14 bands/60 subbands in the fourth ({Tetradecad}). 

The {Spherical Top Data System (STDS)} ({Wenger \& Champion 1998}) 
based on the tetrahedral formalism allows simulations of the line
systems up to the Tetradecad. Subsequent analyses of experimental data
over the past decade have significantly improved the input parameters
for the predictions of the polyads up to the Octad ({Hilico et al.\
2001}). We have used the 2001 version of the STDS software to calculate
line opacity predictions for all available systems, extrapolating up
to upper-state rotational levels of $J\,=\,40$. 
\begin{table}
\vspace*{-1ex}
\caption{STDS line statistics vs.\ 1996 {HITRAN} and {GEISA} data.}
\begin{center}
\begin{tabular}{lclr} \tableline
Transition & $J_{\sf max}$ & $\lambda$[$\,\mu$m] & Number of lines \\
 \tableline
GS $\to$ GS & 40  & 22.2\,--\,10$^6$ &       5470 \\     
GS $\to$ P1 & 40  & 4.59\,--\,15.7 &      57\,441 \\     
GS $\to$ P2 & 40  & 2.56\,--\,5.80 &     218\,243 \\     
GS $\to$ P3 & 40  & 1.77\,--\,3.58 &     631\,728 \\     
GS $\to$ P4 & 40  & 1.35\,--\,2.59 &  1\,608\,020 \\     
P1 $\to$ [P1\,--\,P4] & 25  & 1.79\,--\,10$^6$ &  1\,178\,216  \\
P2 $\to$ [P2\,--\,P4] & 20  & 2.39\,--\,10$^6$ &  6\,593\,824  \\
HITRAN      & 10\,? & 1.6\,--\,100   &      30\,049 \\ 
GEISA       & ?   & 5.0\,--\,10    &         6659 \\ 
\tableline \tableline
\end{tabular}
\vspace*{-1.6ex}
\end{center}
\end{table}
\section{Model Spectra}
Atmosphere models using the new opacity data have been computed with
{PHOENIX} version 12, with other opacity sources included as prescribed
in the AMES-cond grid ({Allard et al.\ 2001}). 
The dust opacities are treated in the limit of complete settling, thus
describing the complete removal of all grains from the photosphere.  
We have tested the models
on the recent observations of {Geballe et al.\ (2002)} and present here
comparisons with the spectrum of the T\,4.5 brown dwarf SDSS\,0207+00,
taken at $\sim\,10\,${\AA} resolution with {NIRCAM/Keck\,II}. 
\begin{figure}
\plotfiddle{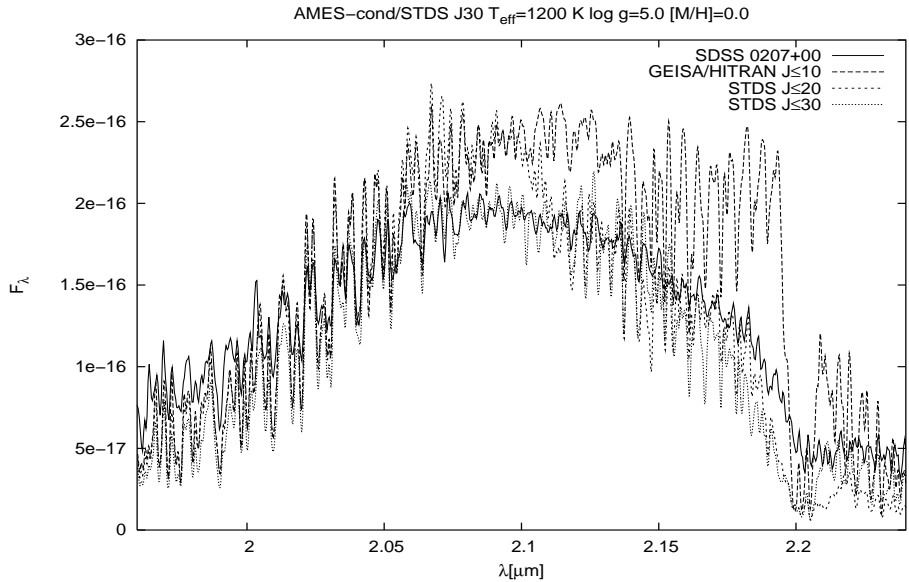}{6.9cm}{0}{98}{88}{-232}{-52}
\caption{Comparison of the K band spectra for the original AMES-cond
model and models using the STDS line opacity predictions to an
observed T dwarf spectrum. The new models include GS-Octad transitions
computed up to $J$ of 20 and 30, respectively.}
\end{figure}

The original AMES-Cond models show a clear lack of opacity in these
regions, particularly on the blue sides of the bands. 
Comparison of the K-band spectra with a preliminary model including
levels up to $J\le 20$ illustrate the improvement achieved by the
additional levels in the Octad system, 
but only the line list including at least levels up to $J\le 30$ gives a
good fit over the full band shape.  

The H band features are more difficult to reproduce with the
STDS predictions, since this wavelength region is dominated by 
transitions to the Tetradecad and higher vibrational levels. 
The current STDS version only includes parameters for the Tetradecad
extrapolated from the analysis of lower polyad systems, which resulted
in a lack of opacity at many wavelengths. 
Newer calculations using preliminary results from the analysis of the
Tetradecad system ({V.\ Boudon 2002, {\em priv.\ comm.}}) 
show considerable improvement over the earlier versions, although
individual line strengths are still not very well reproduced. 

\begin{figure}
\plotfiddle{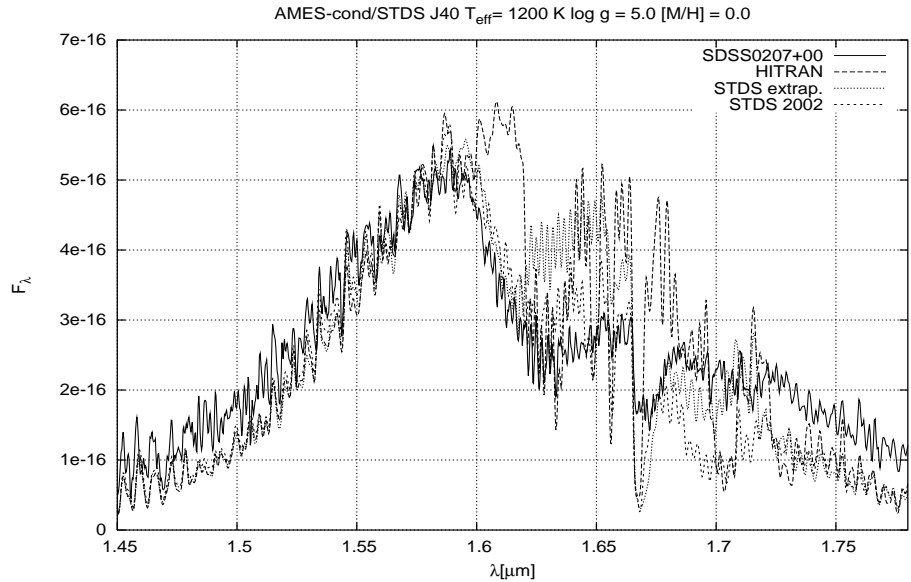}{6.9cm}{0}{98}{88}{-232}{-52}
\caption{H band spectra for AMES-cond model and two STDS models using
extrapolated data from the Octad, and preliminary results from the
Tetradecad analysis.}
\end{figure}

All models still tend to overestimate the absorption in
the regions of strongest opacity, for both CH$_4$ and H$_2$O bands. 
This is probably due to errors in the atmospheric temperature
structure of the current models describing only the limiting case of
full dust condensation. 
This problem will be addressed by a self-consistent model for 
gravitational settling of grains (Allard et al., {\em in prep.}). 

\section{Hot Bands}
 Inclusion of hot bands, transitions originating from higher
polyads, was limited by the upper polyads STDS can handle. 
Thus for the K-band only an extrapolation for the Tetradecad-Dyad was
available, while in the H-band region no hot bands could be computed
at all. Although the upper polyads are sparsely populated at T dwarf
temperatures, due to the steeply increasing degeneracy their
cumulative strength can still contribute a significant background
opacity which might only be treated in a statistical approach (Borysow
et al., 
\href{http://esoads.eso.org/cgi-bin/nph-bib_query?bibcode=2002sam..workE...8B&db_key=AST}{these 
proceedings}). 
These limitations in the hot band models
result in an uncertainy in the total absorption strength that is
comparable in effect to an uncertainty in $T_{\sf eff}$ of
$\sim\,$100\,K or 0.5 dex in log\,$g$ and metallicity.

\section{Conclusions}
STDS-created line lists produce opacity data that significantly
improve on the HITRAN and GEISA databases and allow a much more
detailed analysis of the IR spectra and comparison with
observationally derived spectral indices for brown dwarfs. 
With respect to the contribution of higher polyads several
uncertainties remain that could affect the quantitative determination of
atmospheric parameters based on molecular absorption features. 
A realistic description of the atmospheric structure will also require
a more sophisticated treatment of dust formation which is currently
under way. 

\subsection{Acknowledgements}
This work is supported by NFS grant N-Stars RR185-258, and 
based in part on calculations performed at the NERSC IBM SP with
support from the DoE. We thank V.\ Boudon and J.-P.\ Champion for
helpful information and S.\ Leggett and A.\ Burgasser for access to their
observational data.

\end{document}